\documentclass[aps,twocolumn,superscriptaddress,nofootinbib]{revtex4}

\usepackage{exscale}
\usepackage{graphicx}   
\usepackage{amsmath}
\usepackage{latexsym}
\usepackage{amsfonts}
\usepackage{amssymb}\begin{document}
\title{Quantum switches and quantum memories for matter-wave lattice solitons}

\author {V. Ahufinger}
\affiliation{ICREA and Grup d'\`Optica, Departament de F\'isica, Universitat Aut\`onoma de Barcelona, E-08193 Bellaterra, Barcelona, Spain.}
\author{A. Mebrahtu}
\affiliation{Institut f\"ur Theoretische Physik, Universit\"at
  Hannover, D-30167 Hannover,Germany}
\author {R. Corbal\'an}
\affiliation{Grup d'\`Optica, Departament de F\'isica, Universitat Aut\`onoma de Barcelona, E-08193 Bellaterra, Barcelona, Spain.}  
\author {A. Sanpera}
\affiliation{ICREA and Grup de F\'isica Te\`orica, Departament de F\'isica, Universitat Aut\`onoma de Barcelona, E-08193 Bellaterra, Barcelona, Spain.}

\begin{abstract}
We study the possibility of implementing a quantum switch and a quantum memory for matter wave lattice solitons by 
making them interact 
with "effective"  potentials (barrier/well) corresponding to defects of the optical lattice.
In the case of interaction with an "effective" potential barrier, 
the bright lattice soliton 
experiences an abrupt transition from complete transmission to complete reflection (quantum switch) for a critical height of the barrier. 
The trapping of the soliton in an "effective" potential 
well and its release on demand, without
loses, shows the feasibility of using the system as a quantum memory. 
The inclusion of defects as a way of controlling the interactions between two solitons is also reported. 
\end{abstract}


\maketitle

\section{Introduction}
\label{sec:1}
Bose Einstein condensates (BEC) in optical lattices have attracted during the last years a lot of attention both 
in the mean field regime \cite{meanfield} as well as in the strongly correlated regime \cite{strongly}. 
One of the main reasons for this activity burst is the high level of control achieved in the experiments of ultracold gases in
optical lattices which permits to explore a broad range of fundamental phenomena.

In the mean field regime, a huge interest has been devoted to nonlinear dynamics of matter-waves in periodic media and 
specifically to matter-wave solitons. Matter-wave solitons are self stabilized coherent atomic structures that appear in nonlinear 
systems due to the balance between the nonlinearity and the dispersive effects. The nature of the solitons supported by Bose Einstein 
condensates (BEC) is determined by the character of the interactions: attractive (repulsive) 
nonlinearity supports bright \cite{bright-soliton} (dark \cite{dark-soliton}) solitons. 
In the presence of an optical lattice, this scenario changes completely due 
to the appearance of a band structure in the spectrum and the possibility of having either bright or dark 
lattice solitons with either repulsive or attractive interactions arises. 
Very recently, the first experimental demonstration of bright lattice solitons in repulsive condensates was reported \cite{Oberthaler}.

Since the first proposals of BEC lattice solitons \cite{lattfirst}, there has been an explosion of 
contributions regarding generation, mobility and interactions of this novel type of matter-wave 
solitons both in one-dimensional systems \cite{oned,nosaltres1,nosaltres2} and in higher dimensions \cite{mesd}. 
The interest is mainly centered in bright matter-wave lattice solitons due to their potential applications 
in energy and information transport in nonlinear systems.
The fact that matter wave solitons are massive permits to generate them at rest and to move them
after an appropriate transfer of momentum.
Proposals for controlling the dynamics of bright gap solitons 
are mainly devoted to the manipulation of the optical lattice \cite{control1} and to the modification of the nonlinearity \cite{control2}. 
Nevertheless, a complete control on the dynamics of bright matter-wave gap solitons also requires a profound knowledge of their interactions with defects. 

The interaction of solitons with local inhomogeneities is a subject that appears in the literature 
in different contexts and has been studied in the framework of different nonlinear equations (see for instance \cite{review}). 
In particular, the nonlinear Schr\"odinger equation with point-like defects either in the continuum regime \cite{continu} 
or in discrete systems \cite{discrete} has deserved special attention. Extended defects have also been addressed in this framework \cite{extended}.  
In nonlinear optics, the coupled mode equations have been used to study collisions of moving Bragg solitons with finite size (\cite{coupled1},\cite{coupled2}) and point like defects \cite{coupled3}. Very recently, interactions with defects in the context of continuous matter-wave solitons have also been addressed \cite{defdark, defbright}. 

In this paper we focus on bright lattice matter-wave solitons and propose different possibilities of control of their dynamics by 
making them interact with defects of arbitrary amplitude and width. 
Specifically we will show how to change the direction of movement (a complete
bounce back) of the soliton and how it can be stored and retrieve on demand.
In Sec.~\ref{sec:2} the physical system considered and the model used is introduced. 
In Section ~\ref{sec:3} we present the results concerning the interaction of solitons with an "effective" potential barrier, 
where the possibility of implementing a quantum switch arises. Next, in Sec.~\ref{sec:4} the results regarding the interaction with an "effective" potential well will be shown and the use of the system as a quantum memory will be discussed. The possibility of controlling the interactions between two lattice solitons by placing a defect at the interaction point is discussed in Sec.~\ref{sec:5}. We will conclude in Sec.~\ref{sec:6}.

\section{Physical system}
\label{sec:2}
 
We consider a zero temperature $^{87}$Rb condensate confined in a one-dimensional geometry and in the presence of an optical lattice. 
The description of the system is performed within the one dimensional Gross Pitaevskii equation (GPE):
\begin{equation}
i\hbar \frac{d\psi \left( x,t\right)}{dt} = \left(-\frac{\hbar^2}{2m}\triangle+
V \left( x\right)+g\vert \psi\left( x,t\right)\vert^2\right)\psi \left( x,t\right),
\label{GPE}
\end{equation}
where $g=2 \hbar a_s \omega_{t}$, being $a_s$ the $s$-wave scattering length and 
$\omega_{t}$ the radial angular trapping frequency, is the averaged one-dimensional coupling constant. The external potential is given by:
\begin{equation}
V \left( x \right)=\frac{m}{2} {\omega_x}^2 x^2+ V_0\sin^2(\frac{\pi x}{d}),
\end{equation}
which describes both the axial trapping potential, with angular frequency  $\omega_x$, and the optical lattice, with spatial period $d=\lambda/{2\sin(\theta/2)}$, being $\lambda$ the wavelength of the lasers forming the optical lattice and $\theta$ the angle between them. 
The depth of the optical lattice, $V_0$, is given in units of the lattice recoil energy $E_r=\hbar^2 k^2/2m$ where $k=\pi/d$ is the lattice recoil momentum.  

The generation of the bright lattice soliton is performed as it is reported in \cite{nosaltres1} using parameters close to the experimental realizations \cite{Oberthaler}. The procedure is briefly summarized in what follows. The starting point is a $^{87}$Rb condensate ($a_s=5.8$nm, $m=1.45\times 10^{-25}$Kg) of $N=500$ atoms in the presence of a magnetic trap with frequencies $\omega_{t}=715\times2\pi$ Hz and $\omega_x=14\times2\pi$ Hz and an optical lattice, with potential depth $V_0=1E_r$ and period $d=397.5$nm.
 The axial magnetic trap is suddenly turned off and the appropriate phase imprinting, corresponding to phase jumps of $\pi$ in adjacent sites, is performed \cite{nosaltres1}. After the phase imprinting, the system evolves towards a negative mass, self-maintained staggered soliton at rest centered at $x=0$, which contains approximately $35\%$ of the initial atoms ($N=187$) and extends circa $11$ sites. The exceeding atoms are lost by radiation. 

The total energy of the generated bright lattice soliton can be calculated using the energy functional of the GPE (\ref{GPE}) that contains 
the total kinetic ($E^T_k$), interaction ($E_i$) and potential ($E_p$) energies:
 
\begin{equation}
E=\int\left[\frac{\hbar^2}{2m}|\nabla \psi(x)|^2 +\frac{g}{2}|\psi(x)|^4 +V(x)|\psi(\
x)|^2\right]dx .
\label{functional}
\end{equation}

As clearly observed in the numerical solutions of the GPE, the density profile of the bright lattice soliton inside each well 
is shifted with respect to the minimum of the optical lattice. This shift has to be taken into account to properly calculate the
energy. Assuming a constant shift $\delta$, the Ansatz $\psi(x)=A \exp(-(x-x_0)^2/2 \eta^2)\cos(2\pi x/(\lambda'))$,  with $\lambda'=2d+\delta$, $A$ the amplitude and $\eta$ the width of the Gaussian envelope
\cite{nosaltres2}, leads to the following energy functional: 

\begin{eqnarray}
&&E=B \Bigg( \frac{\hbar^2}{m}\left[\frac{ 1+e^{-{k'}^2 \eta^2}\cos(2{k'}x_0)}{2\eta\
^2}+{k'}^2 \right] \nonumber \\
&&+\frac{g|A|^2}{4\sqrt{2}}\left[3+e^{-2 k'^2 \eta^2}\cos(4k'x_0)+4e^{-\frac{k'^2 \eta^2}{2}}\cos(2k'x_0)\right]\nonumber \\
&&+V_0\Big[1+e^{-k'^2 \eta^2}\cos(2k'x_0)-e^{-k^2 \eta^2}\cos(2k x_0) \nonumber \\
&&  -\frac{1}{2}e^{-k_-^2 \eta^2}\cos(2k_- x_0) -\frac{1}{2}e^{-k_+^2 \eta^2}\cos(2k_+ x_0)\Big]
\Bigg).
\label{energies}
\end{eqnarray}
where $B=|A|^2 \sqrt{\pi}\eta/4$, $k'=2\pi/\lambda'$ and $k_{\pm}=k' \pm k$. Fixing $N=187$, $\delta=0.07\mu$m and $x_0=0$ we
 obtain a minimum of Eq. (\ref{energies}) corresponding to $1.31E_r$. Exact numerical integration of Eq. (\ref{functional}) gives a total energy of $1.35E_r$.
This shows the high level of accuracy that the used variational method provides. Also, an strong agreement is found when we evaluate each term of Eq.(\ref{functional}) 
by using the variational method \cite{nosaltres2} (exact integration): $E^T_k=0.85E_r(0.85E_r)$, $E_i=0.12E_r(0.13E_r)$ and $E_p=0.34E_r(0.37E_r)$.
 
By calculating the linear band spectrum of the system  we obtain an energy at the band edge of $1.25E_r$. This value is in 
good agreement with the total energy obtained previously (Eq.(\ref{functional})) without the non linear term.
The linear band spectrum predicts an 
effective mass at the edge of the first Brillouin zone corresponding to $m_{eff}=-0.15m$.       

Once the lattice soliton is created, it is set into motion by applying an instantaneous transfer of momentum at $t=0$. 
It has to be  large enough to overcome the Peierls-Nabarro (PN) barrier \cite{nosaltres1,kivshar93} but sufficiently small  to assure that the soliton 
remains in the region of negative effective mass, i.e., $0.009k\hbar<p<0.2k\hbar$. The soliton starts to move opposite to the direction 
of the kick, manifesting thus its negative effective mass. The concept of the effective mass is used through the paper to give an intuitive explanation of the observed dynamics. Nevertheless, all the results presented in what follows have been obtained, without any approximation, 
by direct integration of Eq. (\ref{GPE}). 

At a certain distance $x_{m}$ of the initial position of the soliton ($x=0$), the lattice potential $V(x)$ is modified 
in the following way:

\begin{equation}
  \left\{ \begin{array}{ll}
      V_0\sin^2(\frac{\pi x}{d})+ V_{m}(1-\frac{(x-x_{m})^2}{2\sigma^2})& \mbox{\small{if $x_m-l/2\leq x\leq x_m+l/2$}};\\
      V_0\large\sin^2(\frac{\pi x}{d})   & \mbox{\small{otherwise}}.\end{array} \right. 
\label{potmod}      
\end{equation}
where  $V_{m}$ can be either positive or negative and $\sigma=6d$ for all the cases. For $ V_{m}<0$, the local decrease of the lattice potential 
corresponds to an "effective" potential barrier for the soliton due to its negative effective mass while if $V_{m}> 0$, i.e. a local increase 
of the potential
acts as an "effective" well for the soliton. For 
the case of an effective barrier, $x_{m}$ is fixed to match exactly with a 
minimum of the optical lattice while in the effective well case, $x_{m}$ corresponds to a maximum of the optical lattice.
We have checked that the results reported in the following do not depend strongly on the specific shape of the potential 
by reproducing them with Gaussian and square potentials for the defect.

To analyse the interaction of the bright lattice soliton with the defect, it is crucial to know the total energy of the soliton 
while it moves. The applied variational Ansatz with
the shift in the periodicity is meaningful only in the static case and cannot be used to study soliton dynamics,
since the soliton is always chirped with respect its center \cite{nosaltres2}. 
Therefore, to study dynamical behaviour one has to rely on numerical simulations. 
We have numerically calculated the contributions to the total energy of the soliton as a function of time. 
Immediately after the kick, the soliton expels atoms and its energy abruptly decreases becoming much smaller than the energy 
that it would need to remain at rest at the edge of the first Brillouin zone.  
In the framework of the linear band theory this would correspond to displace the particle from the edge of the band of the first Brillouin zone  
by changing its quasimomentum. 
To illustrate the dynamics of the system, we consider the case in which we give a kick of $p=0.1k\hbar$ to the generated soliton at rest. 
At $t=0$, just after the kick, the total energy of the soliton is its energy at rest plus the contribution of the transfer of momentum, i.e. 
$E=1.35E_r+(0.1)^2E_r=1.36E_r$. At $t=1$ms, the soliton energy has decreased 
already to $0.96E_r$, the rest of energy has been taken by the expelled atoms. 
A steady state is reached for a soliton energy of $E=0.92E_r$. While moving, some energy is devoted to cross the PN barrier (the soliton configuration changes  
its shape from a configuration centered in one well of the optical lattice to a configuration centered in one maximum and viceversa). This change of
the shape of the soliton is reflected in the out of phase oscillations of the kinetic energy with respect to the potential plus non linear energy in
such a way that the mean value of the energy remains constant. 

\begin{figure}
\includegraphics[width=1.0\linewidth]{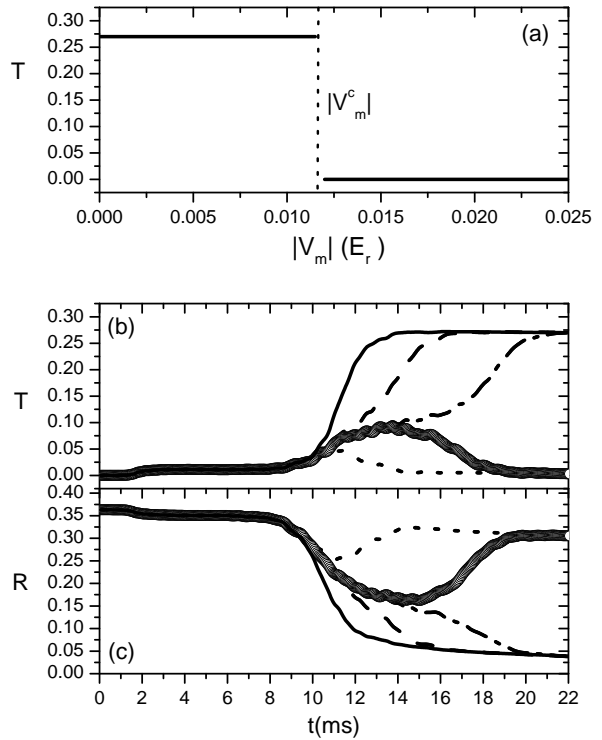}
\caption{(a) Transmission coefficient, $T$, 
as a function of the amplitude of the "effective" potential barrier, $|V_{m}|$; 
(b) Transmission coefficient, $T$ and (c) reflection coefficient $R$ as a function of time for different amplitude defects: $|V_{m}|=0$ (solid line), 
$|V_{m}|=0.01 E_r$ (dashed line), $|V_{m}|=0.0115 E_r$ (dot-dashed line), $|V_{m}|=0.0117E_r$ (circles) and $|V_{m}|=0.018E_r$ (dotted line). 
In all the plots the soliton kinetic energy is $E_k=0.01E_r$ and the width of the defect $l=2d$}\label{barrier1}
\end{figure}

\section{"Effective" potential barrier}
\label{sec:3}

We discuss first the interaction of a bright lattice soliton with an "effective" potential barrier. 
Scattering depends on the width of the defect ($l$) and the relevant energy scale, settled by the ratio $|V_{m}|/E_k$, where $E_k=<P>^2/2m $ is the fraction of the total kinetic energy $E^T_k$ 
devoted to move the soliton, and  $< >$ denotes time average (before reaching the defect). The momentum $P$ is defined as:
\begin{equation}
P(t)=\int -i \hbar \psi^*(x,t) \nabla \psi(x,t) dx.
\label{moment}
\end{equation}
The rest of the kinetic energy is needed to keep the structure and cannot be used to overcome 
the ``effective'' potential barrier.  We have checked that apart from the necessary change in shape to overcome the PN barrier, 
the soliton keeps its overall shape when it reaches the defect. 
This corroborates 
that there is no transfer between non linear energy and kinetic energy apart of the one corresponding to the already discussed PN barrier. 

\begin{figure}
\includegraphics[width=1.0\linewidth]{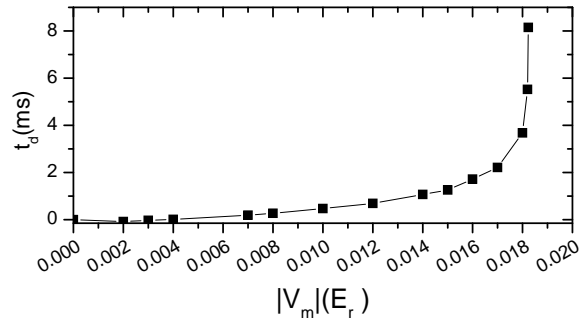}
\caption{Transmission delay time as a function of $|V_{m}|$ for a soliton with kinetic energy 
$E_k=0.01E_r$ crossing a defect of width $l=d$}\label{delay}
\end{figure}

We distinguish two regimes of parameters: (i) when the amplitude of the "effective" potential barrier is on the order of the kinetic 
energy of the soliton ($|V_{m}|\sim E_k$) and (ii) when amplitude of the potential barrier is much larger than the kinetic energy of the soliton. 
In the former case, the potential barrier acts as a quantum switch, {\it i.e.}, either the entire soliton is transmitted or it is completely reflected 
depending on the amplitude of the barrier (Fig. \ref{barrier1}(a)). The transmission ($T$) and reflection ($R$) coefficients are 
calculated by integrating over space (and time) the density of the wavefunction in the region after and before the defect, respectively. Note that since only approximately
35\% of the initial atoms form a soliton and since there are also loses of atoms during the kick, the merit figure for perfect transmission is well below $1$ 
and corresponds approximately to $0.27$ (N=135).  
\begin{figure}
\includegraphics[width=1.0\linewidth]{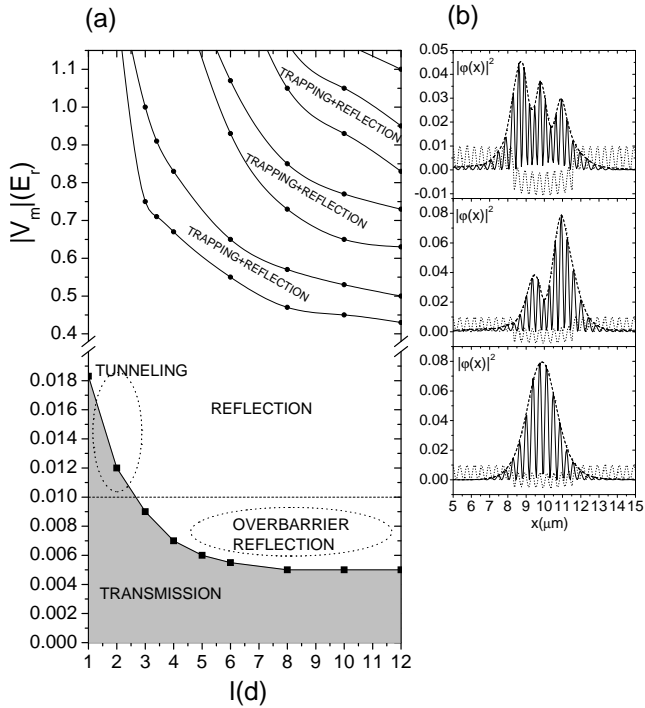}
\caption{(a) Reflection and transmission behavior of a soliton interacting with an effective potential barrier 
as a function of the potential width $l$ (in $d$ units) and potential amplitude $|V_{m}|$ (in $E_r$ units): 
the region of total transmission in gray, and the region of total reflection in white. Inside the reflection region, 
bands in which trapping and reflection occurs appear. 
Region of parameters for which tunneling and overbarrier reflection occur 
are also shown. 
The dashed line shows the value of the kinetic energy of the soliton.
(b) Density profiles of the trapped structure that appears for 
$|V_{m}|=0.55E_r$ (lower plot), $|V_{m}|=0.8E_r$ (middle plot) and $|V_{m}|=1.1E_r$ (upper plot). 
In the three cases $l=8d$.}
\label{diagram}
\end{figure}  
For a fixed width of the defect a drastic change of behavior occurs for a given height of the barrier $|V_{m}^c|$. The wider the defect is 
the lower the critical value of the height of the barrier $|V_{m}^c|$.
The critical values, indicating the transition between complete transmission and complete reflection for different potential widths are shown in Fig.\ref{diagram}(a) 
by solid black squares. Below these values, depicted by a gray region in Fig.\ref{diagram}(a), complete transmission of the soliton occurs. 
As one approaches the critical value from below, the soliton experiences a time delay 
with respect to free propagation (i.e., in the absence of the defect). 
This delay increases as one gets closer to the critical point and eventually the time needed by the soliton to cross the barrier diverges (see Fig. \ref{delay}). 
Above the transition line, reflection of the entire soliton 
occurs after a storage time inside the region of the barrier that also increases as one approaches the critical value. 
To illustrate this behavior, Fig. \ref{barrier1}(b,c) shows transmission and reflection coefficients as a 
function of time for a barrier of fixed width $l=2d$ and  different values of the amplitude.
In the situation shown in Fig. \ref{barrier1}, the critical value is nearly equal to the kinetic energy of the soliton but 
if the width of the barrier is reduced, this critical value can exceed the kinetic energy of the soliton. 
In this case, the soliton tunnels through the barrier, i.e., transmission 
is obtained for values of the amplitude of the barrier higher than the kinetic energy of the soliton (see Fig. \ref{diagram}(a)). 
On the other hand, for wider defects, a region of overbarrier 
reflection appears (Fig. \ref{diagram}(a)). 
There, the lattice soliton is completely reflected 
although it has a kinetic energy larger than the height of the potential barrier. This region extends for a wide range of widths of the defect. 
We have checked that overbarrier reflection occurs even in the limit when the width of the defect is much larger than the size of the soliton. 

Fig. \ref{delay} shows the delay time in transmission with respect to the absence of defect, $t_d$, for $l=d$ as a function 
of the height of the barrier, $|V_{m}|$, including the cases where tunneling occurs ($|V_{m}|> 0.01E_r$). 

Up to now we have described the transition from complete transmission to complete reflection 
by fixing the width of the defect and varying the amplitude. It is worth noticing that a similar switching behavior can be obtained 
by fixing the amplitude of the effective potential and changing its width. This would correspond to horizontal lines in 
Fig. \ref{diagram}(a) crossing the transition line (solid black squares) in the region where $|V_{m}|$ is on the order of the kinetic energy of the soliton. This observed abrupt transition from complete reflection to complete transmission 
opens the possibility to use the system as a quantum switch.  
A similar switching behaviour has been predicted for optical Bragg solitons described with the coupled mode equations \cite{coupled2}. 

Let us now focus on the regime where the amplitude of the barrier is much larger than the kinetic energy of the soliton $|V_{m}|\gg E_k$, 
where the expected behavior is complete reflection of the soliton. Complete reflection occurs 
but there are specific values of the ratio $l/|V_{m}|$ for which the soliton 
splits into two parts: a fraction of the initial soliton becomes trapped inside the region of the barrier while the other part is reflected back 
keeping a solitonic structure. Fig. \ref{diagram}(a), for  $|V_{m}| \gg E_k$, shows the regions where the soliton 
splits into two parts (trapping and reflection) embedded in the complete reflection regions. 
The fraction of atoms trapped inside the defect region has its origin on the atoms lost by radiation due to the repulsive 
force experienced by the soliton when it reaches the potential barrier.
These radiated atoms enter the region of the barrier and for some specific ratios of the width and the height of the defect 
the fraction of trapped atoms increases. These trapping regions appear
as bands as shown in Fig. \ref{diagram}(a).
In each band, the trapped fraction exhibits different spatial distributions: for the first (lowest) one, 
the structure is a single hump; in the second one  a double hump structure appears, and so on (see Fig. \ref{diagram}(b)). 
A noticeable feature of this trapped fraction is that the density maxima of the structure are 
located at the positions of the maxima of the optical potential. 
Increasing the amplitude of the barrier, the structure becomes more independent of the lattice periodicity. 
The extension of the trapped structure is the same independently of the features of the barrier but the number of trapped atoms differs 
for different widths of the barrier. The narrower the defect is 
the larger the number of trapped atoms.
This number changes also with $|V_m|$ inside each band, being maximum at the center of the band.
For all cases the number of atoms forming the reflected soliton 
is always larger than the trapped fraction. 
We have checked that these ``resonance'' bands like do not
correspond to bound states of the linear case. 
We have also observed that this behaviour occurs for all the accessible initial
transfers of momentum that allow motion of the soliton.

\begin{figure}
\includegraphics[width=1.0\linewidth]{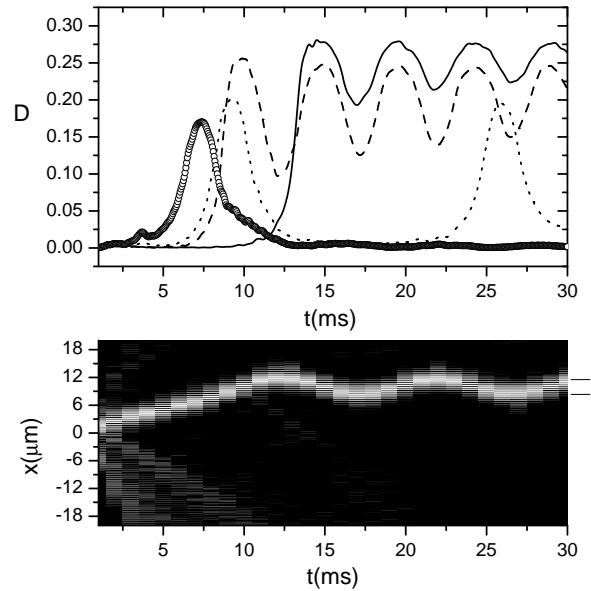}
\caption{(a) Temporal evolution of the trapped fraction of the soliton interacting with an ``effective'' well of depth $|V_{m}|=0.018E_r$ and width $l=8d$
after an instantaneous transfer of momentum of $p=0.05k\hbar$ (solid line), $p=0.1k\hbar$ (dashed line), $p=0.17k\hbar$ (dotted line) and $p=0.2k\hbar$ (circles).
 (b) Contour plot of the evolution in space and time of the lattice soliton with conditions corresponding to the dashed line case in (a).}\label{kinetic}
\end{figure}

\section{"Effective" potential well}
\label{sec:4}

Let us now turn to the interaction of a lattice soliton with an "effective" potential well with a depth of the order of its kinetic energy. For a fixed depth of the well,
 the soliton exhibits different behaviors depending on its kinetic energy. 
For low kinetic energies, the soliton gets bound with the defect and exhibits oscillations
 while for kinetic energies overcoming a certain threshold, the soliton crosses the defect region. In the latter case, the only detectable effect of the potential well
 is the speed up of the soliton with respect to free propagation. It is important to note that as the width of the defect increases, the range of velocities for which 
transmission occurs decreases. To illustrate the described behavior, Fig. \ref{kinetic}(a) shows the time evolution of the trapped fraction density, $D$, for different
 initial transfer of momentum: $p=0.05k\hbar$ (solid line), $p=0.1k\hbar$ (dashed line), $p=0.17k\hbar$ (dotted line) and $p=0.2k\hbar$ (circles) keeping 
the depth ($|V_{m}|=0.018E_r$) and the width ($l=8d$) of the well fixed. The positions in time of the minima of the trapped fraction correspond to the
 turning points of the oscillating 
movement of the soliton around the "effective" well.
The maxima indicates the times for which the soliton is completely inside the well. As expected, the amplitude of the oscillations increases 
with an increasing momentum transfer. 
If the amplitude of the oscillation is larger than the width of the defect, the turning points are located outside the potential well. This is reflected by 
a lower value of $D$. Fig. \ref{kinetic}(b) shows a contour plot of the evolution in space and time of a lattice soliton with $E_k=0.01E_r$ (dashed line case in Fig. \ref{kinetic}(a)). 
The width of the "effective" well is shown at the right hand part
 of the plot to illustrate that indeed the turning points are outside the defect.
\begin{figure}
\includegraphics[width=1.0\linewidth]{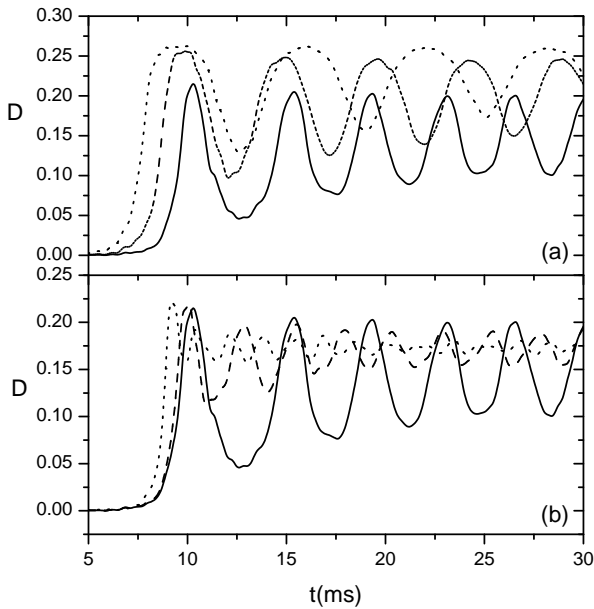}
\caption{Temporal evolution of the trapped fraction density of a lattice soliton with $E_k=0.01E_r$ interacting with an "effective" well 
with (a)$|V_{m}|=0.018E_r$ and $l=4d$ (solid line), $l=8d$ (dashed line) and $l=12d$ (dotted line); (b) $l=4d$ and $|V_{m}|=0.018E_r$ (solid line), 
$|V_{m}|=0.03E_r$ (dashed line) and $|V_{m}|=0.08E_r$ (dotted line)}\label{oscil}
\end{figure}

By fixing $E_k=0.01E_r$, we explore now the dependence of the oscillations on $l$ and $|V_m|$. 
Fig \ref{oscil} (a) displays the temporal evolution of the trapped fraction, 
$D$, for $|V_{m}|=0.018E_r$ and different values of the width of the defect: $l=4d$ 
(solid line), $l=8d$ (dashed line) and $l=12d$ (dotted line). The frequency of the oscillations gives an indication of the width of the defect, decreasing as the width increases, while the amplitude remains approximately the same for all widths. 
In Fig. \ref{oscil}(b) we fix $l=4d$ and display the trapped fraction, $D$, as a function of time for $|V_{m}|=0.018E_r$ (solid line), $|V_{m}|=0.03E_r$ 
(dashed line) and $|V_{m}|=0.08E_r$ (dotted line). 
By inspection of Fig. \ref{oscil}(b), one can confirm that the frequency of the oscillations increases with the depth of the potential while the amplitude 
of the oscillations decreases. This is due to the fact that the soliton experiences a much larger attractive force as the depth of the defect 
increases limiting the displacements around the central position of the well. 

The trapping of the entire lattice soliton around the position of the defect opens possibilities to use the system as a quantum memory because it provides 
the capacity of storage. Nevertheless, in order to perform a memory, one should also be able to release 
the trapped structure after a desirable time and with the minimum loses. We have checked that a soliton trapped in an "effective" potential 
well can be released with a certain velocity keeping the totality of its initial atoms if the defect amplitude is instantaneously set to zero. In fact the 
velocity of the lattice soliton after releasing it will depend on the amplitude of the oscillations it was performing while it was trapped. Specifically, 
the velocity of the structure, after releasing it, grows with the amplitude of the oscillations. Moreover, choosing appropriately the time at which the release takes place, 
one can vary the direction of the movement.      
\begin{figure}
\includegraphics[width=1.0\linewidth]{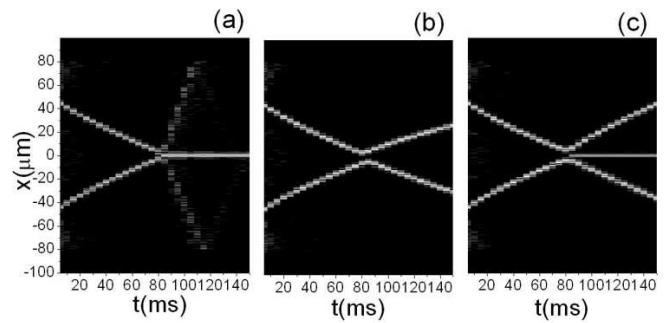}
\caption{Contour plot of the evolution in space and time of a collision between two identical solitons initially placed at symmetrical positions with respect $x=0$ and moving in opposite directions when an "effective" potential barrier with $l=2d$ and (a) $|V_{m}|=0.012E_r$, (b) $|V_{m}|=0.2E_r$ and (c) $|V_{m}|=0.5E_r$ is placed at $x=0$.}\label{colisions}
\end{figure}
\section{Control of the collisions}
\label{sec:5}

Now we investigate if the inclusion of a defect in the lattice helps to control the interactions between two lattice solitons. 
It has been shown that collision of two identical lattice solitons (moving with the same velocity and with the same average phase) 
 merge into a soliton with the same number of atoms as the initial ones \cite{nosaltres2}. The exceeding atoms are lost by radiation. 
If an "effective" potential barrier much narrower than the dimensions of the solitons is placed at the crossing point,
 we find the following behaviours: (i) for $|V_{m}|\leq E_k$ the 
merging behavior is maintained (Fig. \ref{colisions} (a)); (ii) 
for $|V_{m}|\gg E_k$, each soliton reflects back  (Fig. \ref{colisions} (b)). 
Moreover, for some values of $|V_m|$, in addition to the reflection, 
a fraction of atoms trapped in the defect can appear (Fig. \ref{colisions}
(c)). The trapped fraction shows up the same features as in the single soliton case (section III). Modifying the features of the defect, different outcomes
can be engineered. For instance, when the width of the barrier is of the order of the dimensions of the initial solitons, effects like the trapping of both solitons at the edges of the barrier can occur.

\section{Conclusions}
\label{sec:6}

Summarizing, we have found that bright matter wave lattice solitons behave as  ``quantum'' particles when colliding with an ``effective'' barrier/well, 
corresponding to a defect in the optical lattice. Among the rich dynamics 
exhibited by the system, we would like to remark two effects.
The first one corresponds to  the interaction of a soliton with an "effective" potential barrier which permits the implementation of a quantum switch.
In this case, a sharp transition from complete reflection to complete transmission is present at a specific value of the height of the barrier. 
Although this resembles the classical particle behaviour, the quantum nature of the solitons is explicitly manifested in the appearance 
of overbarrier reflection and tunneling.
The second effect we would like to stress appears when the defect acts as 
an "effective" potential well.  We have shown that trapping 
of the entire soliton around the position of the defect 
and its release on demand with a given velocity and direction of motion is possible. This fact indicates the suitability of the system as 
a quantum memory. Finally, it has been also reported 
that the presence of a defect in the lattice 
can help to control the interactions 
of two lattice solitons.

We thank  M. Lewenstein and A. Bramon for fruitful discussions.
We acknowledge support from the Spanish Ministerio de Ciencia y Tecnolog\'ia grants FIS2005-01369, FIS2005-01497 and Consolider Ingenio 2010 CSD2006-00019. A. M. acknowledges financial support from the Deutscher Akademischer Austausch Dienst (DAAD).

\end{document}